\documentclass[aps,prl,twocolumn,noshowpacs,superscriptaddress]{revtex4-1}
\usepackage{graphicx,epsfig, subfigure}
\usepackage{amsbsy}
\usepackage{amssymb}
\usepackage{amsmath}
\usepackage{hyperref}
\usepackage[bottom]{footmisc}
\usepackage{xcolor}
\usepackage{color}
\usepackage{times}
\usepackage{textcomp}
\usepackage{verbatim} 

\date{\today}

\begin{document}

\newcommand{\eqnref}[1]{Eq.~\ref{#1}}
\newcommand{\figref}[2][]{Fig.~\ref{#2}#1}
\newcommand{\RN}[1]{%
  \textup{\uppercase\expandafter{\romannumeral#1}}%
}

\title{Ultrasensitive torque detection with an optically levitated nanorotor}

\author{Jonghoon Ahn}
	\affiliation{School of Electrical and Computer Engineering, Purdue University, West Lafayette, Indiana 47907, USA}
\author{Zhujing Xu}
	\affiliation{Department of Physics and Astronomy, Purdue University, West Lafayette, Indiana 47907, USA}
\author{Jaehoon Bang}
\affiliation{School of Electrical and Computer Engineering, Purdue University, West Lafayette, Indiana 47907, USA}
\author{Peng Ju}
\affiliation{Department of Physics and Astronomy, Purdue University, West Lafayette, Indiana 47907, USA}
\author{Xingyu Gao}
\affiliation{Department of Physics and Astronomy, Purdue University, West Lafayette, Indiana 47907, USA}
\author{Tongcang Li}
	\email{tcli@purdue.edu}
	\affiliation{School of Electrical and Computer Engineering, Purdue University, West Lafayette, Indiana 47907, USA}
	\affiliation{Department of Physics and Astronomy, Purdue University, West Lafayette, Indiana 47907, USA}
	\affiliation{Purdue Quantum Science and Engineering Institute, Purdue University, West Lafayette, Indiana 47907, USA}
	\affiliation{Birck Nanotechnology Center, Purdue University, West Lafayette, Indiana 47907, USA}
	\date{\today}

\begin{abstract}
 Torque sensors such as the torsion balance  enabled the first determination of the gravitational constant by Cavendish \cite{cavendish1798experiments} and the discovery of Coulomb's law. Torque sensors are also widely used in studying small-scale magnetism \cite{wu2017nanocavity,Losby_2018}, the Casimir effect \cite{Chan1941}, and other applications \cite{Hee1600485}. Great effort has been made to improve  the torque detection sensitivity by nanofabrication and cryogenic cooling. The most sensitive nanofabricated torque sensor has achieved a remarkable sensitivity of $10^{-24} \rm{Nm}/\sqrt{\rm{Hz}}$ at millikelvin temperatures in a dilution refrigerator \cite{Kim2016}. Here we dramatically improve the torque detection sensitivity by developing an ultrasensitive torque sensor with an optically levitated nanorotor in vacuum.  We measure a torque as small as $(1.2 \pm 0.5) \times 10^{-27} \rm{Nm}$ in  100 seconds at room temperature. Our system does not require complex nanofabrication or cryogenic cooling.  Moreover, we drive a nanoparticle to rotate at a record high speed beyond 5~GHz (300 billion rpm).   Our calculations show that this system will be able to detect the long-sought  vacuum friction \cite{RevModPhys.71.1233,PhysRevLett.105.113601,Pendry2012,Manjavacas2017} near a surface under realistic conditions. The optically levitated nanorotor will also have applications in studying  nanoscale magnetism \cite{wu2017nanocavity,Losby_2018} and quantum geometric phase \cite{CHEN2019380}. 

\end{abstract}

\maketitle

\begin{figure}[ht]
	\includegraphics[scale=0.4]{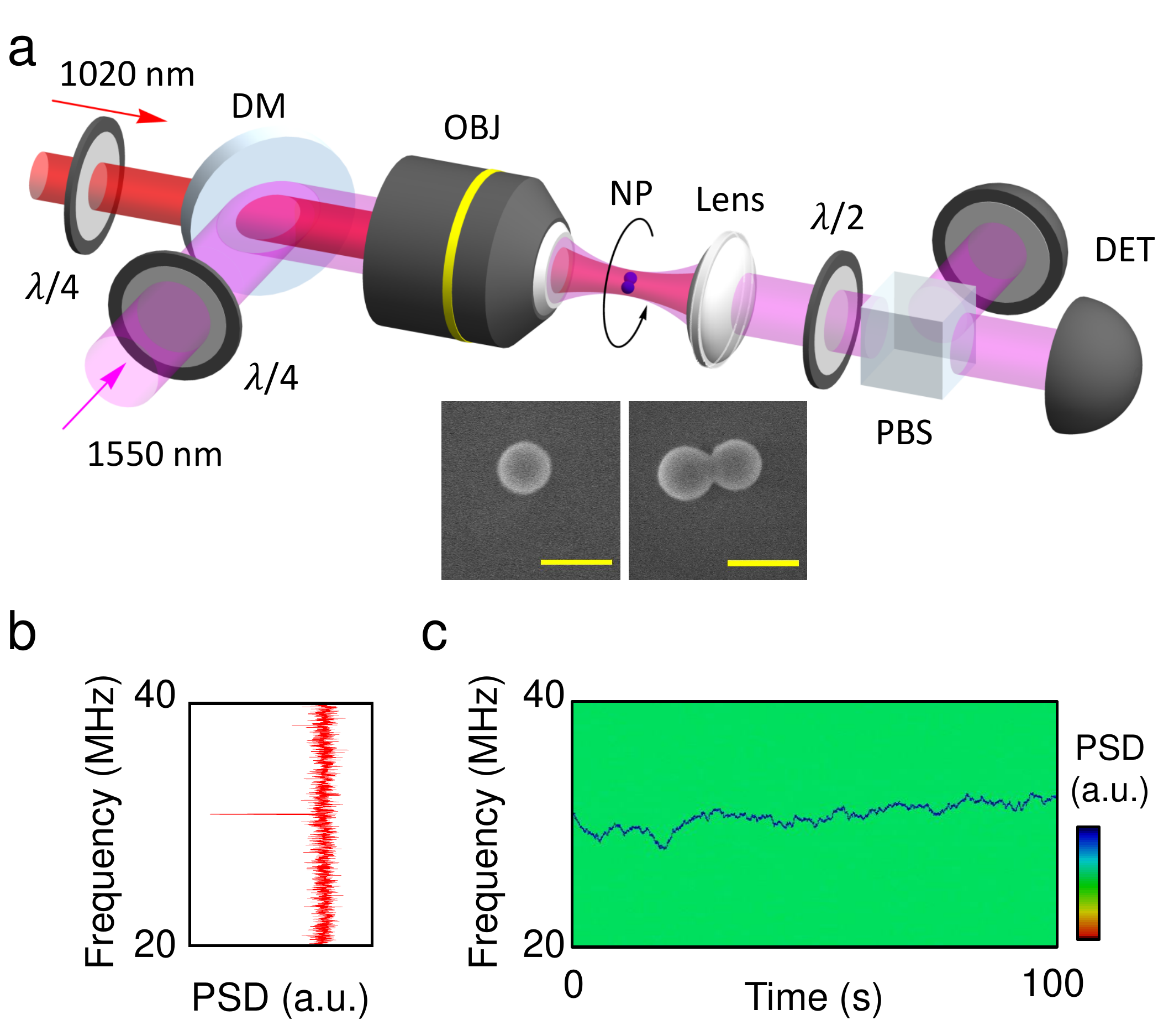}
	\caption{ {\bf Experimental schematic and rotation spectra of an optically levitated nanoparticle.}
	(a) A silica nanoparticle (NP) is levitated in vacuum with a 500 mW, 1550 nm laser tightly focused by an objective lens (OBJ) with NA = 0.85. An additional 1020 nm laser is used to apply an external torque on the nanoparticle.  The polarization of each laser is controlled with a quarter waveplate ($\lambda$/4). After the collimation lens, the trapping laser is directed to detectors for monitoring the motion of the trapped nanoparticle. DM: dichroic mirror;  $\lambda$/2: half waveplate; PBS : polarizing beam splitter; DET: balanced photo detector. Inset: SEM images of a silica nanosphere (left) and a silica nanodumbbell (right). The scale bar is 200 nm for both images. 
	(b) A measured power spectral density (PSD) of the rotation of an optically levitated nanoparticle at $10^{-4}$ torr. The frequency of the PSD peak is twice the rotation frequency of the nanoparticle.
	(c) A spectrogram (time trace) of the rotation PSD of an optically levitated nanoparticle recorded for 100 seconds. The first vertical line corresponds to the PSD shown in (b).  a.u.: arbitrary units.
	}
	\label{scheme}
\end{figure}

Recent developments in levitated optomechanics provide a new paradigm for sensing and precision measurements \cite{Yin2013,Geraci2016,Gratta2016}. Recently, the center-of-mass (COM) motion of an optically levitated nanoparticle in vacuum was cooled to microkelvin temperatures \cite{PhysRevLett.122.223601}.  Experimental control of the  rotation \cite{Arita2013,Kuhn2017,Ahn2018,Reimann2018,Moore2018,Gratta2019}, torsional vibration \cite{Hoang2016,Ahn2018}, and precession \cite{Rashid2018} of a levitated nanoparticle in vacuum have also been demonstrated. A levitated nanoparticle has been used to study nonequilibrium thermodynamics at  small scales \cite{Li1673,Millen2014,Gieseler2014,Hoang2018} and demonstrate force sensing at the zeptonewton scale \cite{Geraci2016}.  It was proposed that an optically levitated nonspherical nanoparticle in vacuum would  be an ultrasensitive  torque sensor \cite{Hoang2016} and could study anisotropic surface interactions \cite{Xu2017}. While optically levitated torque sensors have attracted many interests \cite{Kuhn2017,Rashid2018,Ahn2018}, an experimental demonstration of a torque sensitivity better than that of the state-of-the-art nanofabricated torque sensor ($10^{-24} \rm{Nm}/\sqrt{\rm{Hz}}$) \cite{Kim2016} has not been reported.

\begin{figure*}[ht!]
	\includegraphics[scale=0.45]{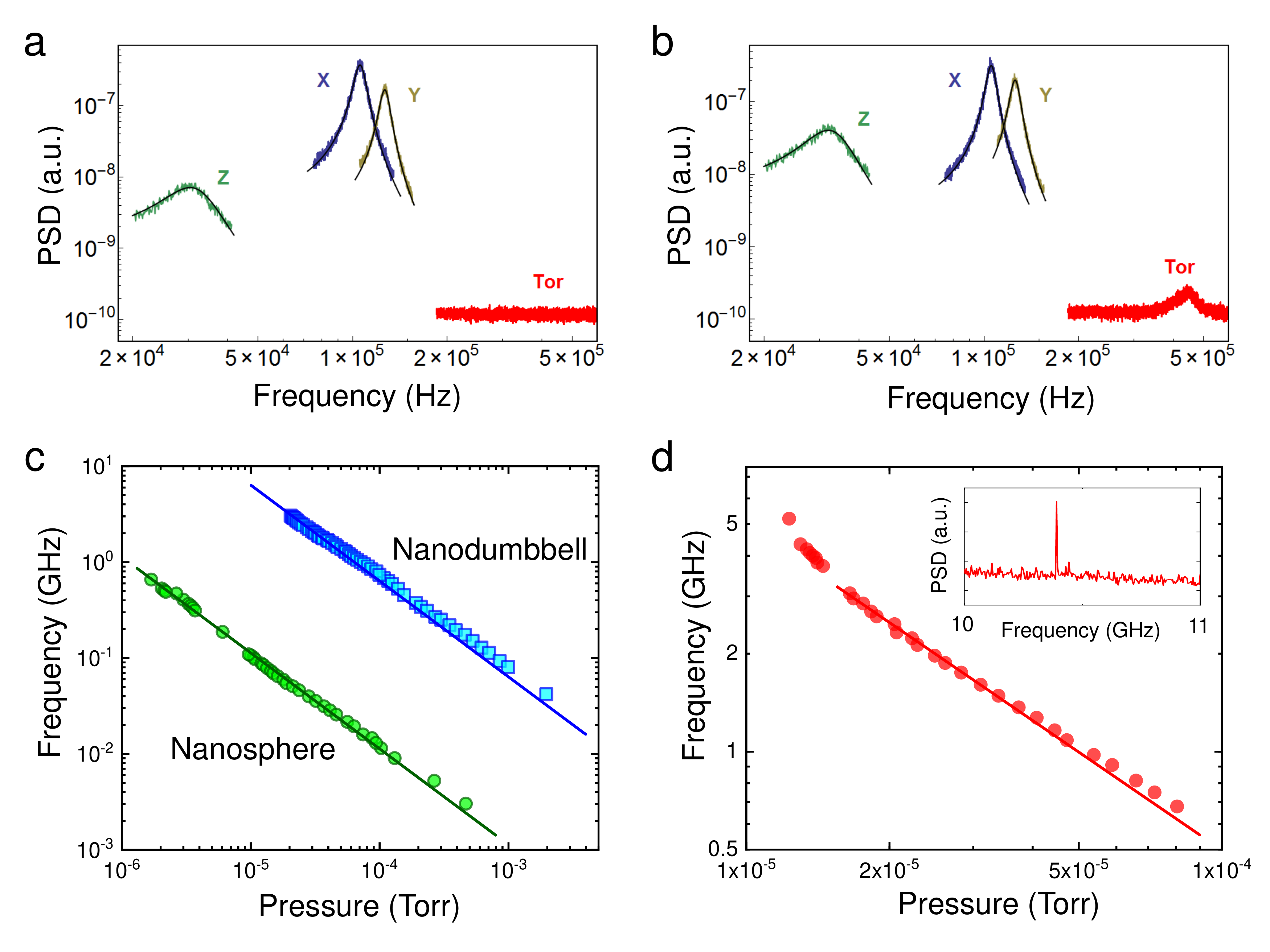}
	\caption{ {\bf Vibration and fast rotation of optically levitated silica nanospheres and nanodumbbells.}
	 PSD's of the motions of a silica nanosphere (a) and a nanodumbbell (b) trapped by a linearly polarized laser at 10 torr. The ratios of the damping rates in directions perpendicular ($y$) and parallel ($x$) to the electric field of the laser  are $1.02 \pm 0.01$ and $1.23 \pm 0.02$ for the nanosphere (a)  and nanodumbbell (b), respectively. Also, an additional torsional peak appears for the nanodumbbell which does not show for the nanosphere. $z$ axis is parallel to the propagating direction of the laser. 
	(c) The rotation frequencies of a nanodumbbell (blue squares) and a silica nanosphere (green circles) in a circularly polarized optical tweezer as a function of the air pressure.  The solid lines show the $1/p$ dependence of the rotation frequencies, where $p$ is the air pressure. The diameters of the nanosphere and  the nanodumbbell are about 150 nm.  (d) Rotation frequency as a function of pressure for a nanoparticle with a large ultimate rotation frequency. The solid line shows the $1/p$ dependence of the rotation frequency. Inset: A measured PSD of the rotational motion. The PSD peak at 10.4 GHz corresponds to a mechanical rotation frequency of 5.2 GHz.
	}
	\label{scheme2}
\end{figure*}

Here we report an optically levitated nanorotor torque sensor that is several orders of magnitudes more sensitive than the state-of-the-art nanofabricated torque sensor \cite{Kim2016}. We measure an external torque as small as $(1.2 \pm 0.5) \times 10^{-27} \rm{Nm}$ in just 100 seconds at room temperature. We also investigate different dynamic behaviors of a nanosphere and a nanodumbbell. 
This nanorotor torque sensor will be particularly suitable to detect the long-sought  vacuum friction \cite{RevModPhys.71.1233,PhysRevLett.105.113601,Pendry2012,Manjavacas2017}. 
A fast rotating neutral nanoparticle can convert quantum and thermal vacuum fluctuations to radiation emission. Because of this, the electromagnetic vacuum behaves like a complex fluid and will exert a frictional torque on a nanorotor \cite{RevModPhys.71.1233,PhysRevLett.105.113601}. While there have been many theoretical investigations on vacuum friction, it has not been observed experimentally yet.  To observe the vacuum friction, the nanorotor needs to spin at a very high speed. In this work, we optically drive a silica nanoparticle to rotate beyond 5 GHz, which is about 5 times faster than the former record of the fastest nanorotor \cite{Ahn2018,Reimann2018}.  We calculate the vacuum friction acting on a rotating silica nanosphere near  a flat surface that has a large local density of electromagnetic states to  enhance the vacuum friction \cite{Pendry2012,Manjavacas2017}.
Our calculations  show that the  vacuum friction acting on a silica nanosphere rotating at 1 GHz near a flat silica surface will be large enough to be observed under realistic conditions.

\begin{figure*}[tb!]
	\includegraphics[scale=0.32]{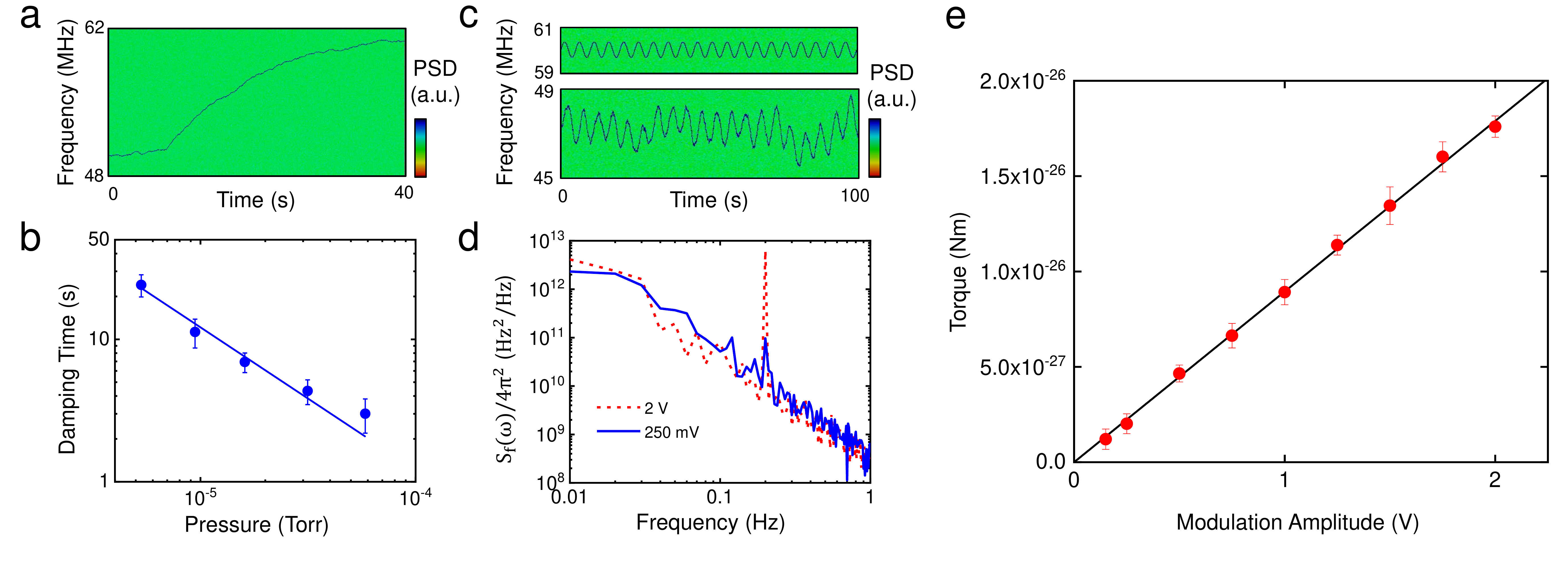}
	\caption{{\bf Ultrasensitive detection of an external torque.}	(a) A spectrogram of the rotation of a trapped silica nanosphere at $9.4 \times 10^{-6}$ torr. After the 1020 nm laser is turned on at 8 s, the rotation frequency increases until its maximum is reached. The frequency trace is fit to an exponential curve to obtain the rotational damping time. (b) The damping time of the rotation as a function of air pressure. The solid line shows $1/p$ dependence. (c) The bottom subfigure shows a rotation spectromgram of the levitated nanosphere under a sinusoidally-modulated external torque. Data was collected at $1.5 \times 10^{-5}$ torr. The frequency and amplitude of the modulation signal that controls the power of the laser is  200 mHz and 2V. 1V corresponds to approximately 75 mW of laser power. For comparison, a RF signal generated by a voltage-controlled oscillator controlled by the same modulation signal is shown in the top subfigure. (d) PSD of the time trace of the rotation frequency. Peaks are due to the modulated external torque. The red dashed line and the blue solid line correspond to a modulation amplitude of 2 V and 250 mV, respectively.  (e) Measured torque for different modulation amplitudes. When the modulation amplitude is 150 mV, the external torque is measured to be  $(1.2 \pm 0.5) \times 10^{-27} $ Nm.
	}
	\label{scheme3}
\end{figure*}

In this experiment, we optically trap a silica nanoparticle (a nanosphere or a nanodumbbell) in a vacuum chamber using a tightly focused 1550 nm laser (Fig. \ref{scheme}).  The polarization of the trapping laser is controlled with a quarter waveplate. An additional 1020 nm laser is used to apply an external torque that will be measured. The  trapping laser passes a collimation lens and is guided to balanced photo detectors that monitor the  rotational, torsional, and the center-of-mass motions of the levitated nanoparticle.  When the nanoparticle rotates, it changes the polarization of the trapping laser slightly, which is monitored with  a balanced photo detector after a polarizing beam splitter (Fig. \ref{scheme}(a)). The signal of the balanced detector is sent to a spectrum analyzer  to measure the rotation frequency.

Once a nanoparticle is trapped in a linearly-polarized 1550~nm laser,  we collect the power spectral density (PSD) signals of its motion at 10 torr to verify its geometry \citep{Ahn2018}. Fig. \ref{scheme2}(a) shows the PSD's of the motion of a nanosphere. The  ratio of the damping rates in directions perpendicular and parallel to the electric field of the laser is measured to be $1.02 \pm 0.01$, which agrees well with the expected value of 1 for a sphere.  There is no observable torsional peak for the nanosphere.  On the other hand, the PSD of a nanodumbbell has a clear torsional peak as shown in Fig. \ref{scheme2}(b). The measured damping ratio is $1.23 \pm 0.02$ for this nanodumbbell, which agrees with the expected value of 1.27 \citep{Ahn2018}. 

 After the geometry of a levitated nanoparticle is confirmed, we change the polarization of the trapping laser from linear to circular. The angular momentum of the circularly polarized laser induces a torque on the levitated nanoparticle and drives it to rotate \cite{Ahn2018,Reimann2018}. The rotation speed is determined by the balance between the optical torque and the drag torque from the surrounding air. Thus the rotation speed is inversely proportional to the air pressure, as shown in Fig. \ref{scheme2}(c). The rotation speed of a nanodumbbell is much faster than that of a nanosphere with the same diameter in the same trap at the same pressure. This is because the optical torque on the nanodummbell is much larger than that on the nanosphere due to  their different shapes. 
Fig.\ref{scheme2}(d) shows the fastest rotation frequency observed in our experiment so far. The rotation rate reaches 5.2 GHz at $1.23 \times 10^{-5}$ torr before the nanoparticle is lost from the optical trap. This is the fastest nanorotor reported to date.

Besides observing the fastest rotation, we employ the nanorotor as an ultrasensitive torque sensor. To test its performance, we use an additional 1020 nm laser to apply an external torque. If we  modulate the 1020 nm laser sinusoidally,   the net torque applied on the nanorotor is 
\begin{equation}
 M_{dc} + M_{ac} \sin(\omega_{m} t) + M_{th} - I \gamma \omega_r= I \dot{\omega_r}  \label{equation1}
\end{equation} 
where $M_{dc}$ is the dc component of the optical torque which mainly comes from the trapping beam, $M_{ac}$ is the external ac torque drive from the 1020 nm laser,  $\omega_{m}$ is the frequency of the modulation, $M_{th}$ is the thermal fluctuation torque, $I$ is the moment of inertia of the nanorotor, $\omega_r$ is the angular rotation velocity of the nanoparticle, $\gamma$ is the rotational damping rate because of residual air molecules. If we ignore the thermal noise $M_{th}$, we  have $\omega_r (t)= \omega_{dc} + \frac {M_{ac}} {\sqrt{I^2 [\omega_{m}^2 + \gamma^2}]} \sin(\omega_{m} t + \phi)$ after the modulated external torque is turned on for a long time. Here  $\omega_{dc}=M_{dc}/(I\gamma)$ is the average rotation frequency and $\phi=\tan^{-1}(-\omega_m/\gamma)$. The rotational damping rate $\gamma$ can be measured experimentally. We can suddenly  turn on the  1020~nm laser and measure the rotation frequency  as a function of time (Fig. \ref{scheme3}(a)). The collected data is fit with an exponential curve $\omega = \omega_1 + (\omega_{2} -\omega_1)(1-e^{-\frac{t-t_1}{\tau}})$. Here, $\omega_1$ is the initial rotation frequency, $\omega_{2}$ is the terminal rotation frequency, $t_1$ is the time when the 1020~nm laser is turned on, and $\tau=1/\gamma$ is the damping time. From the fitting, we determine the damping time.  The measured damping time at different pressures are plotted in Fig.\ref{scheme3}(b). Then  the external AC torque $M_{ac}$ can be measured by observing the change of the rotation frequency $\omega_r(t)$ as a function of time. The sensitivity of measuring an external torque will be limited by the thermal noise torque $M_{th}$. 

Including the effects of the thermal noise, the single-sided power spectrum density  of the time-dependent angular velocity $(\omega_r -\omega_{dc})$ for a measurement time of $\Delta t$ is 
 \begin{equation}
S_f(\omega) 
= \frac{4 k_B T \gamma}{I [\omega^2 + \gamma^2]} + \frac{M_{ac}^2 \Delta t \, {\rm sinc}^2[(\omega - 
\omega_{m}) \Delta t/2]}{2I^2 [\omega^2 + \gamma^2]}.
\label{equation2}
\end{equation} 
 $k_B$ is the Boltzmann constant, and $T$ is the  temperature. Note that  $S_f(\omega)$ can be calculated from the time-dependent rotation frequency $\omega_r/2\pi$ (Fig. \ref{scheme3}(c)) measured by a spectrum analyzer  directly. This is very different from the case of the center-of-mass motion, where a calibration factor is required to convert a measured voltage signal  to the real position \cite{ricci2019accurate}.
  From the $S_f(\omega)$ at the modulation frequency, we can measure the  external AC torque applied to the nanorotor.
Because of the thermal noise, the minimum external torque that can be measured is $M_{min} = \sqrt{\frac{4 k_B T I \gamma}{\Delta t}}$ \cite{Xu2017,Bernard1985}.

We now measure the external torque exerted by the circularly polarized 1020 nm laser according to Eq. \ref{equation2}. We modulate the laser power with a sinusoidal signal at 200 mHz while we measure  the rotational PSD of the nanorotor in real time (bottom subfigure in  Fig.\ref{scheme3}(c)). For reference, we simultaneously monitor a RF signal generated by a voltage-controlled oscillator modulated sinusoidally at the same time  (top subfigure in Fig.\ref{scheme3}(c)). We repeat the measurement for different modulation amplitudes. Each measurement takes 100 seconds. The resulting PSD of the angular velocity $(\omega_r-\omega_0)$ are shown in Fig.\ref{scheme3}(d). We can then calculate the amplitude of the external torque using the PSD and the measured damping time. The minimum resolvable torque  corresponds to the value obtained with no modulation. This is about $6 \times 10^{-28} \, {\rm Nm}$ for a measurement time of 100 seconds at $10^{-5}$ torr. This corresponds to a measured sensitivity of $6 \times 10^{-27} {\rm Nm} / \sqrt{\rm Hz}$, which is comparable to the theoretical thermal-noise limited sensitivity of $3 \times 10^{-27} {\rm Nm} / \sqrt{\rm Hz}$ at  $10^{-5}$ torr at 300K.  The measured sensitivity is several orders improved compared to the state-of-the-art nanofabricated torque sensor in a cryogenic environment \cite{Kim2016}.  As shown in Fig.\ref{scheme3}(e),  we measured an external torque as small as $(1.2 \pm 0.5) \times 10^{-27} \rm{Nm}$ when the modulation amplitude is 0.15 V.

\begin{figure}[tb]
	\includegraphics[scale=0.55]{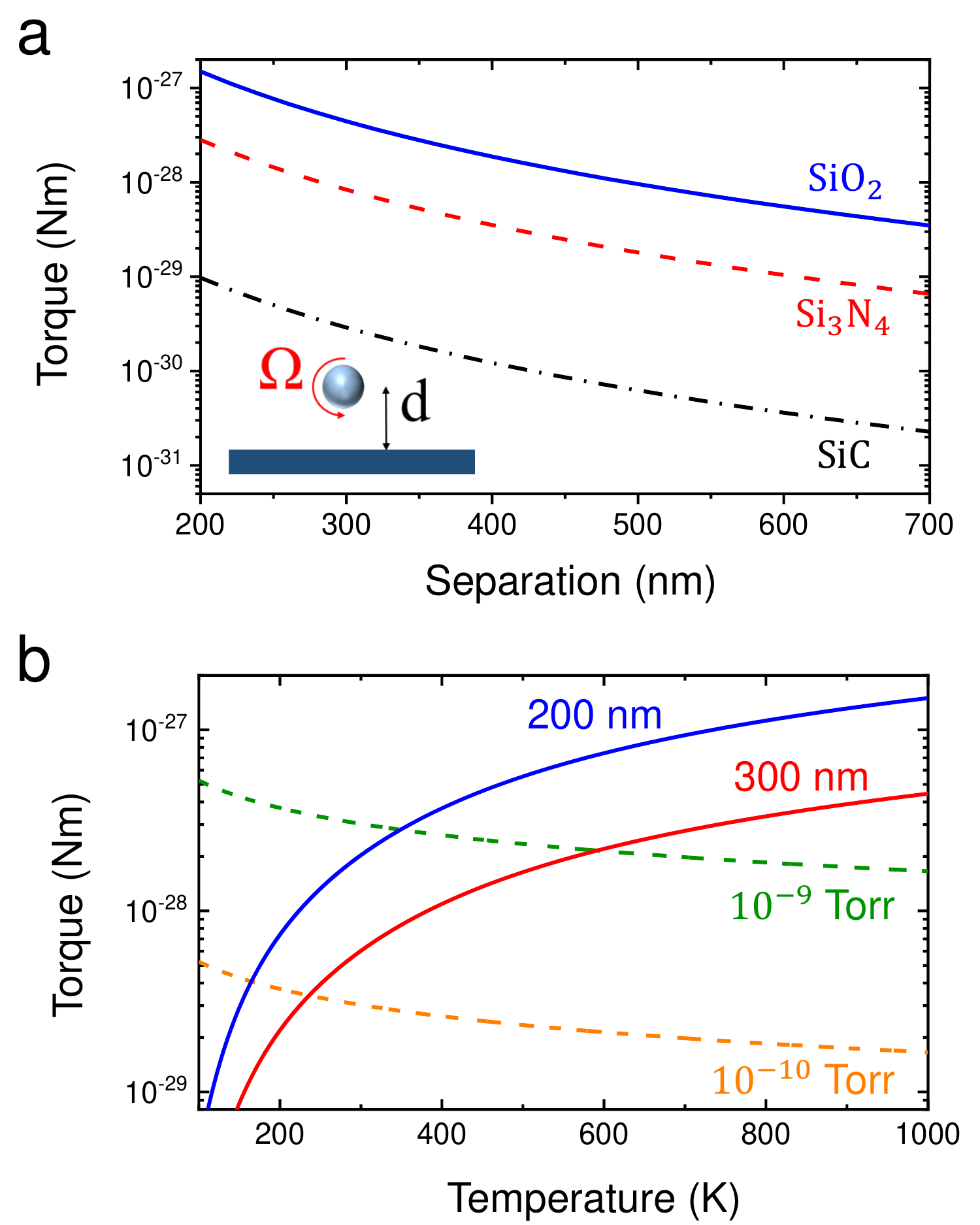}
	\caption{{\bf Calculated vacuum friction on a rotating nanosphere near a surface.}
	(a) Calculated vacuum frictional torque on a rotating silica nanosphere  as a function of the separation between the nanosphere and a substrate. From top down, the curves are for silica (SiO$_2$), Si$_3$N$_4$, and SiC substrates. In the calculations, the radius of the silica nanosphere is 75 nm, its rotation frequency is 1 GHz, and the temperature is  1000 K. Inset: Schematic of a rotating nanosphere  near a substrate. (b) Comparison of the  vacuum frictional torque (solid lines) and the air damping torque (dashed lines)  on a rotating silica nanosphere near a silica substrate as a function of temperature. The blue and red solid curves are  vacuum frictional torques at 200 nm and 300 nm separations, respectively.  The green and orange dashed curves are air damping torques at  $10^{-9}$ torr and $10^{-10}$ torr, respectively. Other parameters are the same as those for (a).
	}
	\label{scheme4}
\end{figure}

One important application of our ultrasensitive  torque sensor with an optically levitated nanorotor will be to detect the long-sought  vacuum friction \cite{RevModPhys.71.1233,PhysRevLett.105.113601,Pendry2012,Manjavacas2017}. Quantum and thermal vacuum fluctuations will create instantaneous charges on the surface of an otherwise neutral nanosphere. If the nanosphere rotates at a high speed, these instantaneous charges will generate radiation, which causes a noncontact friction.
The vacuum friction is  extremely weak in free space \cite{PhysRevLett.105.113601}, but can be enhanced by a nearby surface with a large local density of electromagnetic states \cite{Pendry2012,Manjavacas2017}. 
We perform numerical calculations to find the suitable conditions to detect the vacuum friction. Assuming that a nanosphere is located at a distance $d$ from the surface and rotates at an angular velocity of $\Omega$ around an axis parallel to the substrate, the  vacuum frictional torque is \cite{Pendry2012,Manjavacas2017}:
 \begin{align}
M_{vac} = & - \frac{2 \hbar}{\pi} \int_{-\infty}^\infty[n_1(\omega - \Omega) - n_0(\omega)] \nonumber \\
& \times \operatorname{Im} [\alpha(\omega-\Omega)]\operatorname{Im} [\bar{G}(\omega)] d\omega
\label{equation3}
\end{align} 
where $n_j(\omega) =[\exp(\hbar \omega / k_B T_j)-1]^{-1}$ is the Bose-Einstein distribution function at  temperature $T_j$. For simplicity, we assume the temperatures of the nanosphere ($T_1$) and the substrate ($T_0$) are the same in our calculation. $\alpha (\omega)$ is the electrical polarizability of the nanosphere. $\bar{G}(\omega)=[G_{xx}(\omega)+G_{yy}(\omega)]/2$, where $G_{xx}$ and $G_{yy}$ are electromagnetic Green tensor components \cite{Pendry2012}. $z$ is the axis of rotation. From Eq. \ref{equation3}, we can find that the imaginary part of the polarizability contributes to the vacuum friction. 

We calculate the vacuum frictional torque on a 75 nm-radius silica nanosphere near three different substrates (silica (SiO$_2$), Si$_3$N$_4$, and SiC) at different separations and temperatures (Fig. \ref{scheme4}). These materials support phonon polaritons. Their dielectric functions can be descibed by the Drude-Lorentz model \cite{Pendry2012,Manjavacas2017,Kischkat2012}.  For rotation frequencies much smaller than $k_B T_j/\hbar$, $M_{vac}$ is proportional to the rotation frequency. Inspired by our experimental results, we assume the rotation frequency to be 1 GHz in the calculation. As shown in Fig. \ref{scheme4}(a), a silica surface will give the largest vacuum friction for a rotating silica nanosphere because their phonon polariton modes match. The vacuum frictional torque can be close to $10^{-27}$ Nm at small separations, which is comparable to what we have measured in this experiment. Smaller torques can be measured at lower pressures and for longer times. We also calculated the air damping torque on a rotating nanosphere due to residual air molecules in the vacuum chamber at different pressures. The vacuum friction increases when the temperature increases, while the air damping torque decreases when the temperature increases if the air presure is constant (Fig.  \ref{scheme4}(b)). At $10^{-9}$ torr, the  vacuum frictional torque is larger than the air damping torque when the temperature of the subrate and the nanosphere is larger than 350K at 200 nm separation, or larger than 590 K at 300 nm separation. These temperatures should be easy to achieve. Similar pressures have also been achieved in levitation experiments \cite{Slezak2018, PhysRevLett.122.223601}. Therefore, the detection of the vacuum friction with an optically levitated nanorotor torque sensor is feasible under realistic conditions.

In conclusion, we have demonstrated an ultrasensitive torque sensor with an optically levitated nanorotor in vacuum. We  measure a record small torque of $(1.2 \pm 0.5) \times 10^{-27} \rm{Nm}$ and achieve a record high rotational speed exceeding 5~GHz for a nanorotor. The measured torque sensitivity of our system at room temperature is several orders better than that of the state-of-the-art nanofabricated torque sensor at mK temperatures \cite{Kim2016}. Our system will be suitable to detect the quantum vacuum friction \cite{RevModPhys.71.1233,PhysRevLett.105.113601,Pendry2012,Manjavacas2017}. If the rotating nanoparticle contains an electron spin (e.g. a diamond nitrogen-vacancy center), it can study the quantum geometric phase \cite{CHEN2019380}. 
It can also study nanoscale magnetism, especially the Einstein-de Haas effect and the Barnett effect \cite{Losby_2018}. 

\section*{Acknowledgments}
We thank helpful discussions with F. Robicheaux, T. Seberson, R. Zhao, Z. Jacob, Q. Han, and R. M. Ma. We are grateful to supports from the Office of Naval Research under grant No. N00014-18-1-2371, the NSF under grant No. PHY-1555035, and the Defense Advanced Research Projects Agency (DARPA).


%

\end{document}